\documentclass[aps,prl,twocolumn,showpacs,preprintnumbers,superscriptaddress]{revtex4}
\usepackage{bm,amsmath,amssymb,latexsym,mathrsfs,enumerate,color}
\usepackage[mathcal]{euscript}
\usepackage{graphicx}
\usepackage{hyperref}
\graphicspath{{figs/}}
\renewcommand{\vec}[1]{\bm{#1}}
\begin{document}

\title{Switching of the vortex polarity in a magnetic nanodisk by a DC current}

\author{Denis D. Sheka}
 \email[Corresponding author. Electronic address:\\]{denis\_sheka@univ.kiev.ua}
 \affiliation{National Taras Shevchenko University of Kiev, 03127 Kiev, Ukraine}

\author{Yuri Gaididei}
 \affiliation{Institute for Theoretical Physics, 03143 Kiev, Ukraine}

\author{Franz G.~Mertens}
 \affiliation{Physics Institute, University of Bayreuth, 95440 Bayreuth, Germany}

\date{June 21, 2007}

%
%

\begin{abstract}
We study the dynamics of a vortex state nanodisk due to a dc spin current,
perpendicular to the disk plane. The irreversible switching of the vortex
polarity takes place above some threshold current. The detailed description of
these processes is obtained by spin-lattice simulations.
\end{abstract}




\maketitle

The spin torque effect, which is the change of magnetization due to the
interaction with an electrical current, was predicted by \citet{Slonczewski96}
and \citet{Berger96} in \citeyear{Slonczewski96}. During the last decade this
effect was tested in different magnetic systems
\cite{Tsoi98,Myers99,Krivorotov05} and nowadays it plays an important role in
spintronics \cite{Tserkovnyak05,Marrows05}. Recently the spin torque effect
was observed in vortex state nanoparticles. In particular, circular vortex
motion can be excited by an AC \cite{Kasai06} or a DC \cite{Pribiag07}
spin-polarized current. Very recently it was predicted theoretically
\cite{Caputo07} and observed experimentally \cite{Yamada07} that the vortex
polarity can be controlled using a spin-polarized current. This opens up the
possibility of realizing electrically controlled magnetic devices, changing
the direction of modern spintronics \cite{Cowburn07}.

It was shown in \cite{Caputo07} that in easy-plane Heisenberg magnets a
spin-current which flows perpendicular to the nanoparticle plane acts as an
effective DC magnetic field making energetically unfavorable one of vortex
polarity states. In this Letter we study the magnetic vortex dynamics in
nanodots excited by the spin-polarized current, using the pillar structure,
described in detail in Refs.~\cite{Kent04,Caputo07}. We show that the dipolar
interaction crucially changes the physical picture of the vortex dynamics,
breaking the axial symmetry of the system, i.e. the z-component of the
momentum can be not conserved. Qualitatively speaking the dipolar interaction
makes two main effects \cite{Caputo07a}: (i) there appears an effective
uniaxial anisotropy of the easy-plane type, which is caused by the faces
surface magnetostatic charges and (ii) there appears a nonhomogeneous
effective in-plane anisotropy, which is caused by the edge surface charges
(surface anisotropy). Due to the surface anisotropy the magnetization near the
disk edge is constrained to be tangential to the boundary, which prevents its
precession near the edge. That is why a simple picture of rotational vortex,
which perfectly works for the Heisenberg magnet \cite{Caputo07} should be
revised for the nanodot with account of the dipolar interaction.

The magnetic energy of nanodots consists of two parts: Heisenberg exchange and
dipolar interactions \cite{Akhiezer68}:
\begin{align} \label{eq:Hamiltonian}
\mathcal{H} = &-\frac{\ell}{2}\!\! \sum_{\left(\vec{n},\vec{\delta}\right)}\!
\vec{S}_{\vec{n}}\cdot \vec{S}_{\vec{n}+\vec{\delta}}\\
& + \frac{1}{8\pi}\!\! \sum_{\substack{\vec{n}, \vec{m}\\\vec{n} \neq \vec{m}}}\!
\frac{\vec{S}_{\vec{n}}\cdot \vec{S}_{\vec{m}}-3
\left(\vec{S}_{\vec{n}}\cdot \vec{e}_{\vec{n}\vec{m}} \right)
\left(\vec{S}_{\vec{m}}\cdot \vec{e}_{\vec{n}\vec{m}} \right)}{|\vec{n}- \vec{m}|^3}.
\nonumber
\end{align}
Here $\vec{S}_{\vec{n}}$ is a unit vector which determines the spin direction
at the lattice point $\vec{n}$, $\ell = \sqrt{A/(\mu_0 M_S^2)}$ is the
exchange length ($A$ is the exchange constant, $\mu_0$ is the vacuum
permeability, $M_S$ is the saturation magnetization), the vector $\delta$
connects nearest neighbors, and $\vec{e}_{\vec{m}\vec{n}} \equiv (\vec{n} -
\vec{m})/|\vec{n} - \vec{m}|$ is a unit vector. The lattice constant is chosen
as a unity length.

It is known that as a result of competition between the exchange interaction
and the dipolar one the ground state of a thin magnetically soft nanodisk
is a vortex state: the magnetization lies in the disk plane XY in the main
part of the disk being parallel to the disk edge, forming the magnetic
flux-closure pattern characterized by the vorticity $q=+1$. At the disk center
the magnetization distribution forms a vortex core, which is oriented either
parallel or antiparallel to the z-axis. The former is characterized by a
polarity $p=+1$ and the latter $p=-1$. When electrical current is injected in
the pillar structure, perpendicular to the nanodisk plane, it influences
locally the spin $\vec{S}_{\vec{n}}$ of the lattice through the spin torque
\cite{Slonczewski96,Berger96}
\begin{equation} \label{eq:T}
\vec{T}_{\vec{n}}= j\,\sigma\mathcal{A}
\frac{ \vec{S}_{\vec{n}} \times \bigl[ \vec{S}_{\vec{n}} \times
\hat{\vec{z}}\bigr]}{1+\sigma \mathcal{B} \vec{S}_n\cdot \hat{\vec{z}}}.
\end{equation}
Here $j=J_e/J_p$ is a normalized spin current, $J_e$ is the electrical current
density, $J_p=\mu_0 M_S^2|e|d/\hslash$, $d$ is the disk thickness, $e$ is the
electron charge, $\mathcal{A} =
4\eta^{3/2}/\left[3(1+\eta)^3-16\eta^{3/2}\right]$, $\mathcal{B} =
(1+\eta)^3/\left[3(1+\eta)^3-16\eta^{3/2}\right]$, and $\eta \in(0;1)$ denotes
the degree of the spin polarization; $\sigma=\pm 1$ gives two directions of
spin-current polarization.

\begin{figure*}
\includegraphics[width=\textwidth]{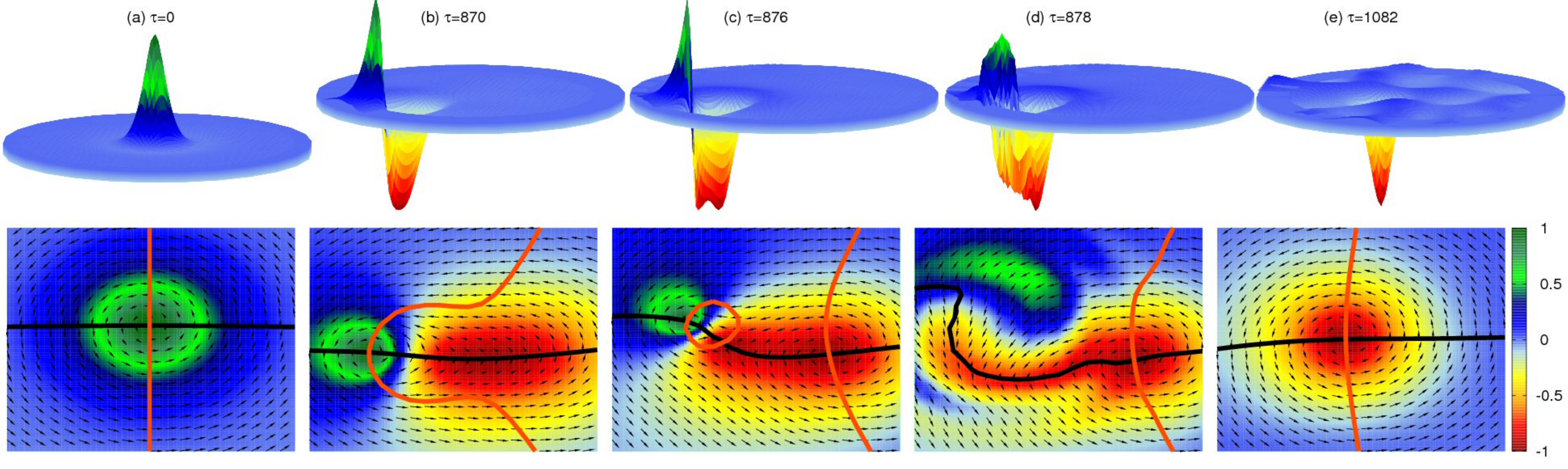}
\caption{(Color online) The time evolution of the vortex switching process
from simulations: the top row shows the 3D distribution of the magnetization
$z$-component, the bottom row corresponds to the in-plane magnetization
distribution near the vortex core. Isosurfaces $S_x=0$ (black curve) and
$S_y=0$ (orange curve) are plotted to determine the vortex position. The current
$j=-0.1$.}
\label{fig:switching}%
\end{figure*}

The spin dynamics of the system is described by the modified
Landau--Lifshitz--Gilbert equation
\begin{equation} \label{eq:LLS-discrete}
\dot{\vec{S}_{\vec{n}}} = -\vec{S}_{\vec{n}}\times \frac{\partial
\mathcal{H}}{\partial \vec{S}_{\vec{n}}} - \alpha \vec{S}_{\vec{n}}
\times \dot{\vec{S}}_{\vec{n}} + \vec{T}_{\vec{n}}.
\end{equation}
Here the overdot indicates the derivative with respect to the dimensionless
time $\tau=\omega_0 t$ with $\omega_0 = 4\pi\gamma M_S$, $\alpha\ll1$ is a
damping coefficient, and $\mathcal{H}$ is the Hamiltonian
\eqref{eq:Hamiltonian}.

\begin{figure*}
\includegraphics[width=\textwidth]{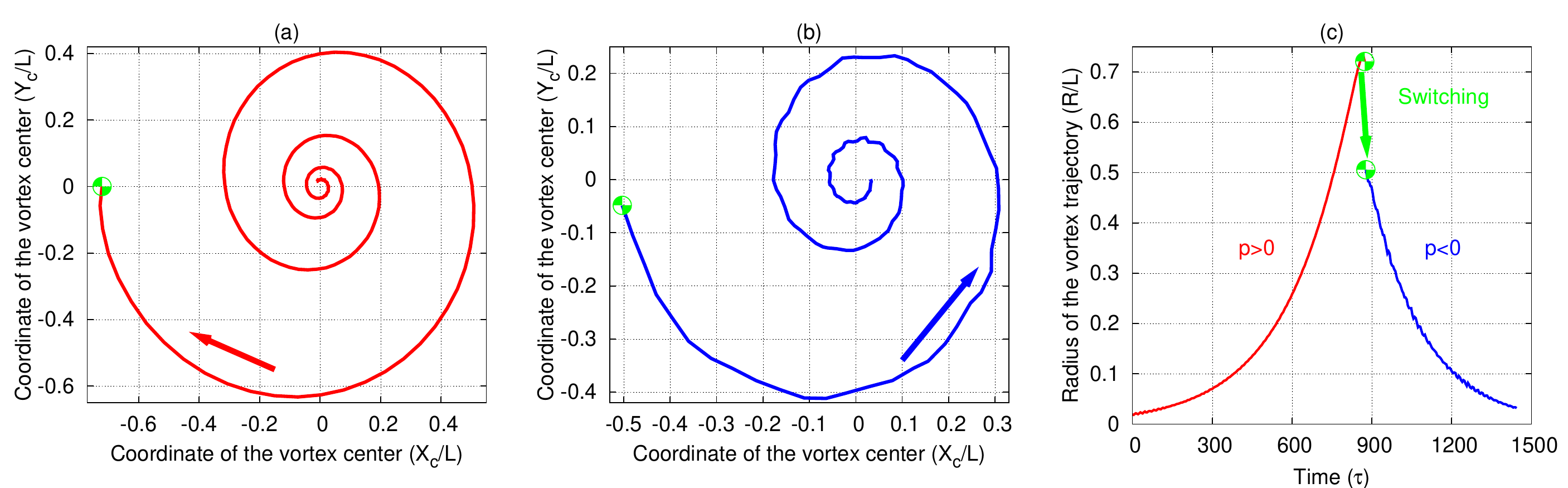}
\caption{(Color online). The vortex dynamics for $j=-0.1$. The vortex
trajectory before the switching (a) and after it (b). The radius of the vortex
trajectory as a function of time (c): at $\tau=877$ the vortex polarity is
switched. } \label{fig:traj}
\end{figure*}

To study the vortex dynamics, we have performed numerical simulations of the
discrete spin-lattice Eq.~\eqref{eq:LLS-discrete}. We consider the case of
thin nanodots where the magnetization does not depend on the z-coordinate. We
have integrated numerically the set of Eqs.~\eqref{eq:LLS-discrete} over the
square lattice using the fourth-order Runge-Kutta scheme with the time step
$\Delta\tau = 0.01$ and free boundary conditions. The lattice is bounded by a
circle of diameter $2L$. In most of the simulations $2L=100$, $h=10$, $\ell =
2.65 $, $\alpha =0.01$, $\sigma=+1$ and $\eta=0.26$.

As an initial condition we use the vortex centered in the disk origin, see
Fig.~\ref{fig:switching}(a). To identify precisely the vortex position we use,
similar to Ref.~\cite{Hertel06}, the crossection of isosurfaces $S_x=0$ and
$S_y=0$, see the bottom row of Fig.~\ref{fig:switching}. The vortex dynamics
results from the force balance between a driving force (by the current), a
dipolar force, a gyroscopical force, and a dissipative force. In the
simulations we observe that when the vortex and the spin-current have the same
polarization ($j\sigma p>0$) the vortex does not quit the center of the disk.
However, for $j\sigma p<0$ ($p=+1$, $\sigma=+1$ and $j<0$ in our case) the
vortex under the action of the current starts to move out of the disk center
following a spiral trajectory, see Fig.~\ref{fig:traj}(a). The spiral type of
motion is caused by the gyroscopical force, which acts on a moving vortex
perpendicular to its velocity in the same way as a Lorentz force acts on a
charged particle in a magnetic field. The role of the charge plays a $\pi_2$
topological charge $Q=qp/2$. The sign of $Q$ determines the direction of the
vortex motion, which is clockwise for $p=+1$. At some point (marked on
Figs.~\ref{fig:traj} by the green symbol) the vortex switches its polarity
($p=-1$). As is seen from Fig.~\ref{fig:switching}, the mechanism of the
vortex switching is very similar to the one observed in other systems
\cite{Waeyenberge06,Xiao06,Hertel07,Lee07,Yamada07,Kravchuk07b}. The moving
vortex excites a non-symmetric magnon mode with a dip situated towards the
disk center. When the vortex moves away from the center, the amplitude of the
dip increases, see Fig.~\ref{fig:switching}(b). When the depth of the dip
reaches a minimum ($S_z=-1$), a pair of a new vortex and antivortex is
created, see Fig.~\ref{fig:switching}(c). The reason why the new-born vortex
tears off his partner has a topological origin. The gyroscopic force depends
on the the total topological charge $Q$. Therefore it produces a
\emph{clockwise} motion for the original vortex $(q=1,p=1,Q=1/2)$ and the
new-born antivortex $(q=-1,p=-1,Q=1/2)$ while the new-born vortex
$(q=1,p=-1,Q=-1/2)$ moves in the \emph{anti-clockwise} direction. As a result
the new vortex separates from the vortex-antivortex pair and rapidly moves to
the origin, see Figs.~\ref{fig:switching}(e) and \ref{fig:traj}(b).
The attractive force between the original vortex $(q=1)$ and the antivortex
$(q=-1)$ facilitates a binding and subsequent annihilation of the
vortex-antivortex pair, see Fig.~\ref{fig:switching}(d).

\begin{figure}
\includegraphics[width=\columnwidth]{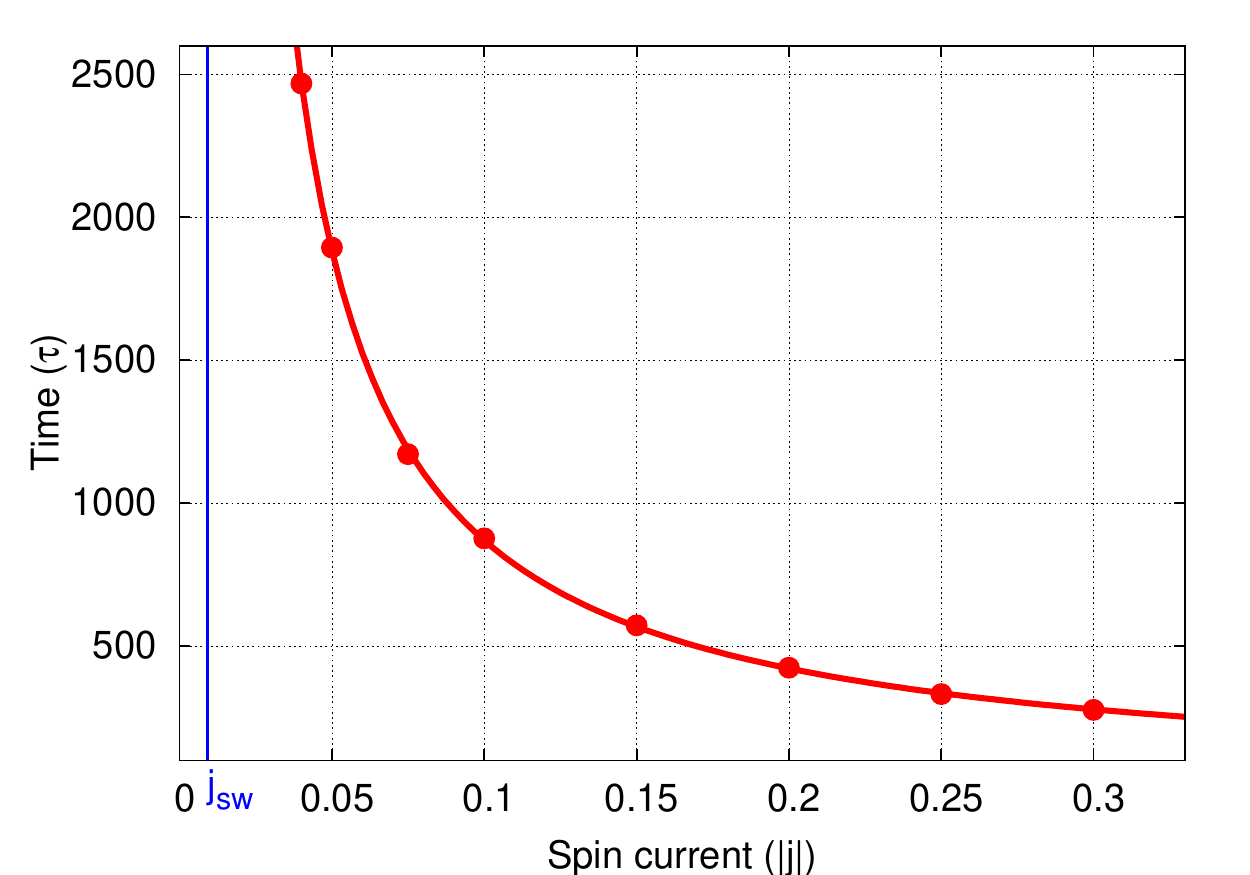}
\caption{(Color online). Switching time as a function of the
applied current.}
\label{fig:sw_time}
\end{figure}

The switching process has a threshold behavior. It occurs when the current
$|j|>j_{\text{sw}}$, which is about $0.012$ in our simulations, see
Fig.~\ref{fig:sw_time}. For stronger currents the switching time rapidly
decreases. Using typical parameters for permalloy disks
\cite{Caputo07,Kravchuk07b} ($\eta = 0.26$, $A=26$ pJ/m, $M_S=860$ kA/m,
$\alpha = 0.01$), we estimate that the time unit $1/\omega_0=50$ ps, the
critical current density is about 0.1 A/$\mu$m$^2$ for a nanodot of 20nm
thickness. The total current for a disk with diameter 200 nm is about $10$~mA.

To summarize, we have studied the magnetic vortex switching under the action
of a DC electrical current. We showed that the switching mechanism is
essentially the same as in the cases when it is induced by a magnetic field
pulse \cite{Waeyenberge06,Xiao06,Hertel07}, by an AC oscillating \cite{Lee07}
or rotating field \cite{Kravchuk07b}, or by an in-plane electrical current
\cite{Yamada07}. There are two key points in this process: (i) the dipolar
interaction causes the deformation of the magnetization profile for the fast
moving vortex, which finally results in the creation of an additional
vortex-antivortex pair, (ii) the topological charge structure of these three
excitations secures the survival of the vortex with the new polarity direction
or, in other words, the polarity switching. The detailed study of the vortex
dynamics including the switching process is under construction.

\acknowledgments

The authors thank S.~Komineas and V.~Kravchuk for helpful discussions. D.S.,
Yu.G. thank the University of Bayreuth, where this work was performed, for
kind hospitality and acknowledge the support from DLR grant No.~UKR~05/055.
D.S. acknowledges the support from the Alexander von Humboldt--Foundation.


\end{document}